# All-optical image denoising using a diffractive visual processor

Çağatay Işıl[1,2,3], Tianyi Gan[1,3], F. Onuralp Ardic[1], Koray Mentesoglu[1], Jagrit Digani[1], Huseyin Karaca[1], Hanlong Chen[1,2,3], Jingxi Li[1,2,3], Deniz Mengu[1,2,3], Mona Jarrahi[1,3], Kaan Akşit[4], and Aydogan Ozcan[1,2,3,*]

[1]Electrical and Computer Engineering Department, University of California, Los Angeles, CA, 90095, USA

[2]Bioengineering Department, University of California, Los Angeles, CA, 90095, USA

[3]California NanoSystems Institute (CNSI), University of California, Los Angeles, CA, 90095, USA

[4]University College London, Department of Computer Science, London, United Kingdom

* ozcan@ucla.edu

**Abstract**

Image denoising, one of the essential inverse problems, targets to remove noise/artifacts from input images. In general, digital image denoising algorithms, executed on computers, present latency due to several iterations implemented in, e.g., graphics processing units (GPUs). While deep learning-enabled methods can operate non-iteratively, they also introduce latency and impose a significant computational burden, leading to increased power consumption. Here, we introduce an analog diffractive image denoiser to all-optically and non-iteratively clean various forms of noise and artifacts from input images – implemented at the speed of light propagation within a thin diffractive visual processor that axially spans $< 250 \times \lambda$, where $\lambda$ is the wavelength of light. This all-optical image denoiser comprises passive transmissive layers optimized using deep learning to physically scatter the optical modes that represent various noise features, causing them to miss the output image Field-of-View (FoV) while retaining the object features of interest. Our results show that these diffractive denoisers can efficiently remove salt and pepper noise and image rendering-related spatial artifacts from input phase or intensity images while achieving an output power efficiency of ~30-40%. We experimentally demonstrated the effectiveness of this analog denoiser architecture using a 3D-printed diffractive visual processor operating at the terahertz spectrum. Owing to their speed, power-efficiency, and minimal computational overhead, all-optical diffractive denoisers can be transformative for various image display and projection systems, including, e.g., holographic displays.

**Introduction**

Image denoising is a fundamental problem encountered in various fields, such as computational imaging and displays [1], computer vision [2], and computer graphics [3]. For example, in computational imaging, noise removal from images is used to mitigate the effects of various sources of noise and artifacts, e.g., image sensors, channel transmission, and environmental conditions [4], [5]. Similarly, within the realm of computer graphics, image denoising is crucial for reducing the low-sampling related image rendering artifacts frequently encountered in real-time graphics processing applications [6]–[8].

Over the past several decades, numerous algorithms have been developed for image noise removal. [9]–[11]. Apart from his renowned contributions to the birth of holography, Dennis Gabor proposed one of the earliest methods for image denoising, involving the Gaussian smoothing of noisy images [12]. A plethora of other algorithms emerged for image denoising, including, e.g., Wiener filtering [2], anisotropic filtering [13], total variation (TV) denoising [14], denoising by soft-thresholding [15], bilateral filtering [16], non-local means denoising [17], block-matching and 3D filtering (BM3D) [18], and among many others (*8–10*). While quite powerful, these classical denoising techniques often need many iterations for their inference, making them less suitable for real-time applications. Deep Neural Networks (DNNs) have also significantly impacted the field of image denoising, especially in the last decade [19], [20]. These artificial DNNs have many parameters that are stochastically optimized (trained) using supervised learning with a large number of noisy-clean image pairs forming the training image set. After their training, DNNs generally operate in a non-iterative feed-forward fashion and have shown remarkable performance advantages for image denoising of unknown input images, never seen before [21]–[28]. It was also reported that DNN-based image denoisers could be used for real-time applications, including interactive Monte Carlo rendering [6], [7], [29]–[31]. Despite the recent improvements in modern graphics processing units (GPUs), achieving interactive speeds in Monte Carlo rendering necessitates working with low spatial sampling, resulting in artifacts in the rendered images. DNN-based denoisers have been proposed to mitigate such artifacts for real-time applications, demanding the use of relatively costly and resource-intensive GPUs [32], [33].

Here, we report an analog diffractive image denoiser (Fig. 1) designed to all-optically process noisy phase or intensity images to filter out noise at the speed of light propagation through a thin diffractive visual processor – optimized using deep learning. Our diffractive denoiser framework consists of successive passive modulation layers that are each transmissive; this diffractive architecture forms a coherent image processor that all-optically scatters out the optical modes representing various forms of noise and spatial artifacts at the input images, causing them to miss the output image Field-of-View (FoV), while passing the optical modes representing the desired spatial features of the input objects with minimal loss and aberrations, forming denoised images at the output FoV without any digital computation in its blind inference. Following its one-time supervised learning-based training performed on a computer, a diffractive image denoiser can work at any part of the electromagnetic spectrum by scaling the dimensions of its optimized diffractive features in proportion to the wavelength of light ($\lambda$), eliminating the need to redesign its layers for different wavelengths of operation.

We demonstrate the capabilities of this analog diffractive image denoiser framework on both phase and intensity images, mitigating salt and pepper noise and low-sampling related spatial image artifacts. Our analyses show that these all-optical denoisers successfully filter out various types of image noise or artifacts at the input using a thin diffractive processor that axially spans <250×λ, while achieving a decent output power efficiency of ~30-40%. For a proof-of-concept, the presented diffractive denoiser framework was also experimentally validated at the terahertz spectrum for removing salt-only random noise in intensity input images using 3D-printed diffractive layers optimized via deep learning. This physical image denoiser framework presents a rapid and power-efficient solution for all-optical filtering of image noise or artifacts, and can potentially be used for holographic displays and projectors operating at different parts of the electromagnetic spectrum.

**Results**

In this manuscript, the terms "diffractive visual processor", "diffractive image denoiser", "diffractive optical network", and "all-optical image denoiser" are interchangeably used. Figure 1 illustrates the schematic of two different diffractive image denoisers trained to all-optically filter out salt and pepper noise from noisy phase or intensity input images; the first one of these diffractive image denoisers (Fig. 1a) is trained to perform phase-to-intensity transformations, whereas the second one (Fig. 1b) is trained to perform intensity-to-intensity transformations between the input and output FoVs. A comprehensive analysis of the all-optical image denoising performances of these trained diffractive denoiser designs under various levels of salt and pepper noise is demonstrated in Fig. 2. In these numerical analyses, each one of the all-optical image denoisers has 5 diffractive layers, which were optimized/trained using the *tiny quickdraw* dataset [34]. The input illumination is considered to be a uniform plane wave (monochromatic and spatially coherent), and the noisy input images to be filtered are in the form of either phase-encoded or intensity-encoded images (see Fig. 2a). For each input encoding type (phase/intensity), different diffractive denoisers were trained using noise probabilities ($\boldsymbol{P}_{tr}$) sampled uniformly from $U(0, \rho)$ where $\rho \in \{0.1, 0.2, 0.4\}$; $\boldsymbol{P}_{tr}$ determines the ratio of the image pixels affected by noise relative to the overall pixel count of the image; see the Methods for details. These training noise probabilities ($\boldsymbol{P}_{tr}$) were randomly sampled for each batch of the input images during each epoch of the training, and the noise-free case ($\boldsymbol{P}_{tr} = 0$) in Figs. 2b-c corresponds to our baseline designs trained with input images free from noise or artifacts. All the trained models were blindly tested using the tiny *quickdraw* test dataset for different test noise probabilities ($\boldsymbol{P}_{te}$). Peak-Signal-to-Noise Ratio (PSNR) and Structural Similarity Index Measure (SSIM) were used as image quality metrics to quantify the all-optical denoising performance of the trained models [35]. Further information regarding the architecture of the diffractive image denoisers, the noise models, the training loss functions, the datasets, and other aspects of our implementation are reported in the Methods section.

Figure 2b illustrates the all-optical image denoising results of diffractive denoisers trained for noisy phase-encoded input images (salt and pepper noise). The output intensities shown for two test images illustrate the success of the trained diffractive image denoisers (with $\boldsymbol{P}_{tr} \sim U(0, \rho)$ where $\rho \in \{0.1, 0.2, 0.4\}$) compared to a conventional diffractive imager trained without noise (i.e., $\boldsymbol{P_{tr}} = 0$). For instance, the denoising performance of the diffractive model trained using $\boldsymbol{P_{tr}} \sim U(0, 0.2)$ demonstrates superior performance for phase-encoded test images created with $\boldsymbol{P_{te}} = 0.1, 0.2,$ and $0.4$, achieving average PSNR improvements of 0.65, 1.47, and 1.90 dB, respectively, when compared to the diffractive imager trained without noise, $\boldsymbol{P_{tr}} = 0$. A similar conclusion can be drawn for all-optical filtering of the intensity-encoded noisy images reported in Fig 2c. The diffractive image denoiser trained using $\boldsymbol{P_{tr}} \sim U(0, 0.2)$ exhibits an improved denoising performance when compared to the baseline diffractive imager ($\boldsymbol{P_{tr}} = 0$), achieving average output image PSNR improvements of 0.83, 1.39, and 1.45 dB for different noise levels of $\boldsymbol{P_{te}} = 0.1, 0.2,$ and $0.4$, respectively.

These numerical results reported in Fig. 2 demonstrate the versatility of the all-optical image denoiser framework to filter out salt and pepper noise present at the input phase or intensity images. These diffractive image denoisers effectively learn to filter out the spatial modes that statistically represent the targeted noise features, while successfully transferring the spatial modes representing the desired features of the input objects, forming denoised intensity images at the output FoV with minimal optical loss and aberrations. In this sense, a diffractive image denoiser can be considered a 3D spatial filter composed of successive phase gratings structurally optimized through supervised deep learning to physically couple out undesired spatial modes of targeted noise features, causing them to miss the output image FoV.

In addition to salt and pepper noise, we also designed diffractive image denoisers to mitigate image artifacts stemming from the Monte Carlo-based low-sampling image renderings, as depicted in Fig. 3. In this analysis, we report the results of different diffractive denoisers, which were trained using noise rates ($\boldsymbol{\gamma_{tr}}$) sampled uniformly from $U(0, \rho)$ where $\rho \in \{1, 2, 3\}$ for both phase-encoded and intensity-encoded input images. $\boldsymbol{\gamma_{tr}}$ indicates the noise rate of the Monte Carlo image renderings, as detailed in the Methods section. Diffractive models with $\boldsymbol{\gamma_{tr}} = 0$ define our baseline, trained with noise-free input images. The denoising capabilities of these diffractive models were blindly tested for various levels of test noise, $\boldsymbol{\gamma_{te}} \in \{0, 1, 2, 3\}$; see the Methods section for details. The results of these analyses are reported in Figs. 3b-c, which demonstrate the advantages of all-optical image denoisers for both phase-encoded and intensity-encoded input images, further supporting the conclusions of the earlier analyses in Fig. 2. For example, Fig. 3b present the results of the diffractive models trained using phase-encoded images with $\boldsymbol{\gamma_{tr}} \sim U(0, 3)$, which outperform the diffractive imager trained with $\boldsymbol{\gamma_{tr}} = 0$ for various test images, improving the average PSNR values by 0.39, 3, and 1.84 dB for $\boldsymbol{\gamma_{te}} = 1, 2,$ and $3$, respectively. Similarly, Fig. 3c demonstrates the success of the all-optical diffractive denoiser models trained using intensity-encoded input images, achieving average PSNR improvements of, e.g., 2.22 and 1.45 dB for $\boldsymbol{\gamma_{te}} = 2$ and $3$, respectively, in comparison to the diffractive imager trained using $\boldsymbol{\gamma_{tr}} = 0$.

These results reported in Figs. 2-3 demonstrate the *internal* generalization capabilities of the trained diffractive denoisers as the test images (although never seen before) were acquired from the same dataset (*tiny quickdraw*). To explore the *external* generalization of the all-optical denoisers to different datasets containing images with distinct spatial features, we conducted additional tests using Fashion MNIST and EMNIST image datasets [36], [37], as illustrated in Fig.

4. The trained diffractive image denoiser, which mitigates the low-sampling artifacts of Monte Carlo renderings ($\gamma_{tr} \sim U(0,2)$) from phase-encoded input images, and the baseline diffractive imager trained with $\gamma_{tr} = 0$ are both tested with noisy input images with different levels of noise ($\gamma_{te} \in \{0,1,2\}$). The average PSNR and SSIM values calculated across the corresponding test datasets confirm the external generalization capabilities of the all-optical image denoisers. For example, as illustrated in Fig. 4a, the diffractive image denoiser trained with $\gamma_{tr} \sim U(0,2)$ achieves average PSNR (SSIM) improvements of 1.74 dB (0.157) for $\gamma_{te} = 1$ and 3.37 dB (0.285) for $\gamma_{te} = 2$, when compared to the baseline diffractive imager trained without noise ($\gamma_{tr} = 0$). A similar analysis is reported in Supplementary Fig. S1, which further confirms the external generalization capabilities of diffractive image denoisers trained with salt and pepper noise.

One of the essential characteristics of all-optical image denoisers, as well as other diffractive visual processors, is their output diffraction efficiency. In the previously demonstrated results reported so far, the all-optical image denoising performance of these diffractive models was achieved without employing a training loss term to penalize low diffraction efficiency at the output FoV. To understand the trade-off between the output diffraction efficiency and the image quality, we conducted additional analysis reported in Fig. 5. As detailed in the Methods section, we adjusted the output diffraction efficiency of an image denoiser by varying the weight ($\beta$) of the diffraction efficiency loss term. During the training process of each diffractive image denoiser model, noisy input images (*tiny quickdraw* dataset) were used, after being subjected to salt and pepper noise with a noise probability ($P_{tr}$) sampled uniformly using $U(0,0.2)$. These image denoiser models trained with various $\beta$ values were subsequently tested on images that were affected by salt and pepper noise with a noise probability of $P_{te} = 0.1$. As depicted in Fig. 5a, for the denoising of phase-encoded images, an all-optical diffractive denoiser can achieve ~28% diffraction efficiency with negligible degradation in its output image quality (~0.08 dB and ~0.004 decrease in the average PSNR and SSIM values, respectively). Similarly, for intensity-encoded images, all-optical denoisers can be designed to have up to ~34% diffraction efficiency while incurring a negligible decrease in output image quality (e.g., ~0.11 dB and ~0.016 in the average PSNR and SSIM values, respectively); see Fig. 5b.

For an experimental proof-of-concept of the presented technique, we built a 3-layer diffractive image denoiser that was trained for noisy intensity images with salt-only noise, with the noise probability ($P_{tr}$) uniformly sampled from the interval $U(0,0.2)$. As depicted in Fig. 6a, the resulting diffractive design was then fabricated and precisely aligned for experimental testing using a single-pixel THz detector and a continuous-wave THz illumination source ($\lambda$ = ~0.75 mm). Figure 6b shows sample binary intensity images of a handwritten letter under different levels of salt-only noise, along with their photographs after the fabrication. Furthermore, the phase profiles of the trained layers of the diffractive image denoiser and photographs of these layers after their fabrication through 3D printing are illustrated in Fig. 6c. During the training of this diffractive model, small random 3D misalignments were introduced to the positions of the diffractive layers to ensure a physically resilient design, which is also referred to as the "vaccination" of the diffractive model [38], [39]. The schematic of the experimental setup is also depicted in Fig. 6d; see the Methods section for further details. The trained diffractive image denoiser model was experimentally tested on several binary intensity images with various levels of salt-only noise, determined by different noise probabilities ($P_{te} \in \{0, 0.05, 0.1, 0.15\}$). Figure 7 provides the optical layout and the experimental results of the 3-layer diffractive denoiser using these noisy inputs. We observe a very good concordance between the numerical and experimental results presented in Fig. 7c, validating the accuracy and resilience of the 3D-fabricated all-optical diffractive image denoiser. The success of these measurements provides an experimental proof-of-concept of the presented framework for all-optical image denoising.

**Discussion**

We introduced a deep learning-enabled diffractive image denoiser framework capable of addressing various forms of noise inherent to different input types, e.g., phase or intensity images. In our analyses, the all-optical denoisers were used to remove both salt and pepper noise and the spatial artifacts originating from the Monte Carlo low-sample image renderings that are typically addressed using nonlinear filters and deep neural networks running on GPUs. The presented diffractive image denoisers successfully filter out these different types of noise at the input using analog processing of the input object waves; this process effectively couples out the characteristic modes that statistically represent the noise features using the sub-wavelength phase structures of the diffractive layers optimized through deep learning. These phase structures are also optimized to cause minimal optical loss and aberrations for the traveling waves that represent the characteristic modes of the input objects. In this sense, the diffractive image denoiser can be considered a smart analog spatial mode filter composed of successive phase gratings with a lateral pitch of ~λ/2. Additionally, these all-optical image denoisers do not consume any power during the filtering operation except for the illumination light that diffracts through passive layers. Regarding the output diffraction efficiency, our findings reveal that these analog image denoisers can achieve e.g., ~30-40% power efficiency without significantly compromising their image denoising performance. The presented diffractive image denoising framework was also demonstrated experimentally, using a 3D-printed diffractive model, successfully removing salt-only noise from input images as illustrated in Figs. 6 and 7.

In our analyses and results, the denoising capabilities of the diffractive denoisers have been demonstrated for the salt and pepper noise and the low-sampling artifacts of Monte Carlo image renderings. Especially for real-time imaging applications, the removal of noisy Monte Carlo renderings is a critical challenge, which has led to the development of various deep learning-based digital image denoisers [6], [7], [29]–[31]. Compared to these digital approaches, the all-optical analog operation of diffractive image denoisers enables the processing of input images as the light diffracts through very thin optical elements that axially span <250 λ; this ultra-high speed and power efficiency of diffractive image denoisers would especially be important for real-time image processing applications.

On the other hand, there are also limitations of the presented approach. First, fabricating a multi-layer diffractive visual processor with phase elements densely packed with a lateral feature size of ~λ/2-λ is challenging, especially for visible and IR wavelengths, due to the tight alignment and fabrication requirements. To mitigate these challenges and develop 3D fabrication processes optimized for diffractive network models, there have been various efforts to fabricate 3D diffractive networks operating in the visible and IR wavelengths [40]–[42]. To bring the presented framework to shorter wavelengths in, e.g., the visible band, different methods of 3D nano-fabrication, such as two-photon polymerization and optical lithography, can be used to manufacture and align the resulting diffractive layers of a diffractive image denoiser. In addition to these, vaccination strategies [38], [39] have been introduced to mitigate the impact of fabrication errors and physical misalignments by intentionally (and randomly) introducing such variations to the layers of a diffractive model during the training process to have more robust diffractive systems that can better withstand physical imperfections. One disadvantage of such vaccination efforts is that the independent degrees of freedom within the diffractive processor are reduced since the vaccination process effectively increases the feature size at the diffractive layer; this, however, can be mitigated by using wider and deeper architectures involving, e.g., a larger number of diffractive layers that are each wider.

A second limitation of the presented approach is that we only considered monochromatic illumination that is spatially coherent. While this assumption can be justified for certain applications that utilize, e.g., holographic image projection/display set-ups, it is also possible to extend the design of all-optical image denoisers to operate under spatially and temporally incoherent light. Diffractive optical networks, in general, form diffraction-limited universal linear transformers between an input and output FoV, and can be trained using deep learning to operate at various illumination wavelengths [43]–[48], also covering spatially incoherent illumination [49]. Therefore, all-optical image denoisers and the underlying design framework can be extended to filter/denoise color images (e.g., RGB) or even spatially and temporally incoherent hyperspectral image signals.

In summary, we presented power-efficient and ultra-high speed all-optical image denoisers that filter out input image noise in the analog domain without consuming any power except for the illumination source. The success of all-optical image denoisers can inspire the creation of all-optical visual processors crafted to solve various other inverse problems in imaging and sensing.

**Methods**

**Diffractive image denoiser design**

An all-optical image denoiser contains a series of diffractive surfaces $l = 0,1,...,L-1$, each of which is located at a different axial position $z_l$. The field transmittance of each diffractive surface $T_l(x, y)$ that is used to modulate the coherent wavefield $U_l(x, y)$ is stochastically optimized using deep learning [50]. The modulated coherent wavefield $U'_l(x, y) = U_l(x, y)T_l(x, y)$ is propagated to the axial position of next diffractive layer $z_{l+1}$ using the angular spectrum method, based on the Rayleigh-Sommerfeld diffraction integral that represents a 2D linear convolution of the propagation kernel $w(x, y, z)$ and the modulated wavefield $U'_l(x, y)$:

$$U_{l+1}(x, y) = U'_l(x, y) * w(x, y, z_{l+1} - z_l),$$

$$w(x,y,z) = \frac{z}{r^2}\left(\frac{1}{2\pi r} + \frac{1}{j\lambda}\right)\exp\left(j\frac{2\pi r}{\lambda}\right),$$

$$r = \sqrt{x^2 + y^2 + z^2}$$

(1)

where $U_{l+1}(x,y)$ denotes the coherent wavefield at the axial position $z_{l+1}$. The field transmittance function of each surface $T_l(x,y)$ is defined as:

$$T_l(x,y) = exp\left(j\frac{2\pi}{\lambda}(\tau(\lambda) - n_a)H_l(x,y)\right)$$

(2)

where $\tau(\lambda) = n(\lambda) + j\kappa(\lambda)$ is the complex refractive index of the optical material, $n_a = 1$ refers to the refractive index of the medium (air in our case) surrounding the layers, and $H_l(x,y)$ represents the thickness profile of the corresponding diffractive surface, which is defined as

$$H_l(x,y) = O_l(x,y)(h_m - h_b) + h_b$$

(3)

where $O_l(x,y)$ is an auxiliary variable array used to compute the thickness value for each $(x,y)$ point between $[h_b, h_m]$. $O_l(x,y)$ and consequently the thickness profile $H_l(x,y)$ for each diffractive layer $l$ are jointly optimized using deep learning to obtain field transmittance function $T_l(x,y)$ for each surface [50]–[52].

**Vaccination of the diffractive image denoisers**

To mitigate the impact of potential misalignments in the physical implementation of a diffractive processor, error sources were integrated into the forward model during the training of the diffractive design that was experimentally demonstrated. These error sources are modeled by 3D displacement vectors, $D^l = (D_x, D_y, D_z)$ corresponding to the difference in the position of diffractive layer $l$, from its ideal location, where $D_x, D_y$, and $D_z$ were defined as uniformly distributed random variables,

$$D_x \sim U(-\Delta_x, \Delta_x),$$
$$D_y \sim U(-\Delta_y, \Delta_y),$$
$$D_z \sim U(-\Delta_z, \Delta_z)$$

(4)

where $\Delta_x, \Delta_y,$ and $\Delta_z$ represents the maximum displacements along the $x$-, $y$-, and $z$- axes, respectively. Thus, the position of the diffractive layer $l$ at $i^{th}$ iteration $L^{(l,i)}$ is defined as

$$L^{(l,i)} = (L_x^l, L_y^l, L_z^l) + (D_x^{(l,i)}, D_y^{(l,i)}, D_z^{(l,i)}).$$

(5)

**Training and testing datasets**

In our numerical results, we used 72,000 randomly selected images from the quickdraw dataset [34]. These images (28 x 28 pixels) were augmented by random rotations ($\Theta \sim U(-15°, 15°)$) and padded to 32 x 32 pixels. Then, they were split into three sets of images including 60,000 training, 2,000 validation, and 10,000 test images. The prepared dataset is called *tiny quickdraw dataset*. To analyze the external generalization of the trained models, we also tested the resulting diffractive designs with unseen images from datasets different from the *tiny quickdraw dataset* including 14,400 EMNIST handwritten letters test images (interpolated to 32 x 32 using bicubic kernel) and 10,000 Fashion MNIST test images (scaled by 0.8 and interpolated to 32 x 32) [36], [37].

For the experimentally demonstrated design, the EMNIST dataset was used, which was split into two sets containing 80,000 training and 8,800 validation images. These datasets, along with the EMNIST test dataset (14,400 images) were interpolated to 20 x 20 pixels and used for the optimization and evaluation of experimentally-tested diffractive image denoiser. Without loss of generality, the contrasts of the images were inverted to facilitate the 3D fabrication of noisy objects in our experiments.

**Implementation details of all-optical denoisers for the numerical results**

The smallest feature size of a transmissive diffractive layer and the sampling period of the propagation model were chosen as $0.5\lambda$. Input and output FoVs of the diffractive image denoisers were $16\lambda \times 16\lambda$ (32 x 32 in pixels). In addition, the size of each diffractive layer was chosen as $64\lambda \times 64\lambda$ (128 x 128 in pixels). The window size of the propagation model was defined as 256 x 256 in pixels, and the matrices representing the FoVs and the diffractive layers were padded with zeros to avoid aliasing. In the numerical simulations, the material absorption was assumed to be zero ($\kappa(\lambda) = 0$), which is a valid assumption considering the overall thickness of our diffractive processor, axially spanning $< 250 \times \lambda$. The axial distance between two consecutive planes was chosen as $40\lambda$. The phase coefficient function of each layer $\theta_l(x,y) = \frac{2\pi}{\lambda}(n(\lambda) - n_a)H_l(x,y)$ and consequently the field transmittance function $T_l(x,y)$ were stochastically optimized using deep learning. $\theta_l(x,y)$ were initialized as 0 for each layer.

**Implementation details of the experimental results**

A monochromatic THz illumination source ($\lambda = \sim 0.75\ mm$) was used in the experiments. Input/output FoVs were determined to be $40\lambda \times 40\lambda$ (3 cm × 3 cm) and the size of each diffractive layer was selected as $66.67\lambda \times 66.67\lambda$ (5 cm × 5 cm). The diffractive feature width of the layers and the sampling period of the propagation model were chosen as $\sim 0.667\lambda$. The pixel size at the measurement plane was $\sim 1.33\lambda$, which is equivalent to the noise feature size at the input FoV. To accurately fabricate the transmissive diffractive layers and the noisy/clean input objects, the complex refractive index of the 3D-printing material $\tau(\lambda)$ was measured as $\sim 1.6518 + j0.0612$. During the training of the experimentally-tested diffractive image denoiser, the thickness profile of each trainable layer $H_l(x,y)$ was optimized in the range $[0.5\ mm, \sim 1.65\ mm]$ that corresponds to $[-\pi, \pi)$ for phase modulation. For experimental testing, a 3-layer all-optical image denoiser was trained, fabricated, and tested. The axial distance between the input plane and the first diffractive layer was set to ~13.34$\lambda$. The other axial distances between successive layers were chosen as ~66.67$\lambda$. To have a misalignment-resilient design, the positions of the layers and the object were randomly shifted during training following the vaccination strategy outlined in Eq. 5. The maximum axial and lateral misalignments $\Delta_x$, $\Delta_y$, and $\Delta_z$ were chosen as ~0.26$\lambda$, ~0.26$\lambda$, and ~0.5$\lambda$, respectively. The thickness profiles of the trained diffractive surfaces and noisy/clean input objects were converted into STL files using MATLAB and they were fabricated by using a 3D printer (Objet30 Pro, Stratasys Ltd.).

The schematic diagram of the experimental setup is shown in Fig. 6d. The incident wave was generated using a modular amplifier (Virginia Diode Inc. WR9.0M SGX)/multiplier chain (Virginia Diode Inc. WR4.3x2 WR2.2x2) (AMC) with a compatible diagonal horn antenna (Virginia Diode Inc. WR2.2). An RF input signal of 10dBm at 11.1111 GHz ($f_{RF1}$) generated by a synthesizer (HP 8340B) was used as the input and multiplied 36 times to produce continuous-wave (CW) radiation at 0.4THz. The AMC was modulated with a 1kHz square wave for lock-in detection. The axial distance between the exit aperture of the horn antenna and the object plane of the 3D-printed diffractive image denoiser was ~75cm and the aperture of the horn antenna is measured to be ~4mm × 4mm. The output FoV of the diffractive denoiser was scanned using a 0.25 × 0.5 mm detector with a step size of 0.75mm. To enhance the Signal-to-Noise Ratio (SNR) and better align with the output pixel size of our design, which was ~1.33$\lambda$ (1 mm), a 3×3 bilinear upsampling and 4×4-pixel binning were used. The signals were detected by a Mixer (Virginia Diode Inc. WRI 2.2) equipped with a pinhole (0.25 × 0.5 mm) placed on an XY positioning stage composed of vertically combined two linear motorized stages (Thorlabs NRT100). A 10 dBm RF signal at 11.0833 GHz ($f_{RF2}$) was sent to the detector as a local oscillator to down-convert the signal to 1 GHz for further measurement. The down-converted signal was amplified by a low-noise amplifier (Mini-Circuits ZRL-1150-LN+) and filtered using a 1 GHz (+/-10 MHz) bandpass filter (KL Electronics 3C40-1000/T10-O/O). The signal was initially measured by a low-noise power detector (Mini-Circuits ZX47-60) and read by a lock-in amplifier (Stanford Research SR830) with the 1kHz square wave serving as the reference signal. The raw data were subsequently calibrated into a linear scale.

**Supplementary Information** contains:

- Supplementary Figure S1
- Image noise models
- Training loss function and imaging performance comparison metrics

[53] Y. Quan, M. Chen, T. Pang, and H. Ji, "Self2Self With Dropout: Learning Self-Supervised Denoising From Single Image," presented at the Proceedings of the IEEE/CVF Conference on Computer Vision and Pattern Recognition, 2020, pp. 1890–1898. Accessed: Sep. 15, 2023. [Online]. Available: https://openaccess.thecvf.com/content_CVPR_2020/html/Quan_Self2Self_With_Dropout_Learning_Self-Supervised_Denoising_From_Single_Image_CVPR_2020_paper.html

[54] I. Loshchilov and F. Hutter, "SGDR: Stochastic Gradient Descent with Warm Restarts," presented at the International Conference on Learning Representations, Nov. 2016. Accessed: Sep. 16, 2023. [Online]. Available: https://openreview.net/forum?id=Skq89Scxx

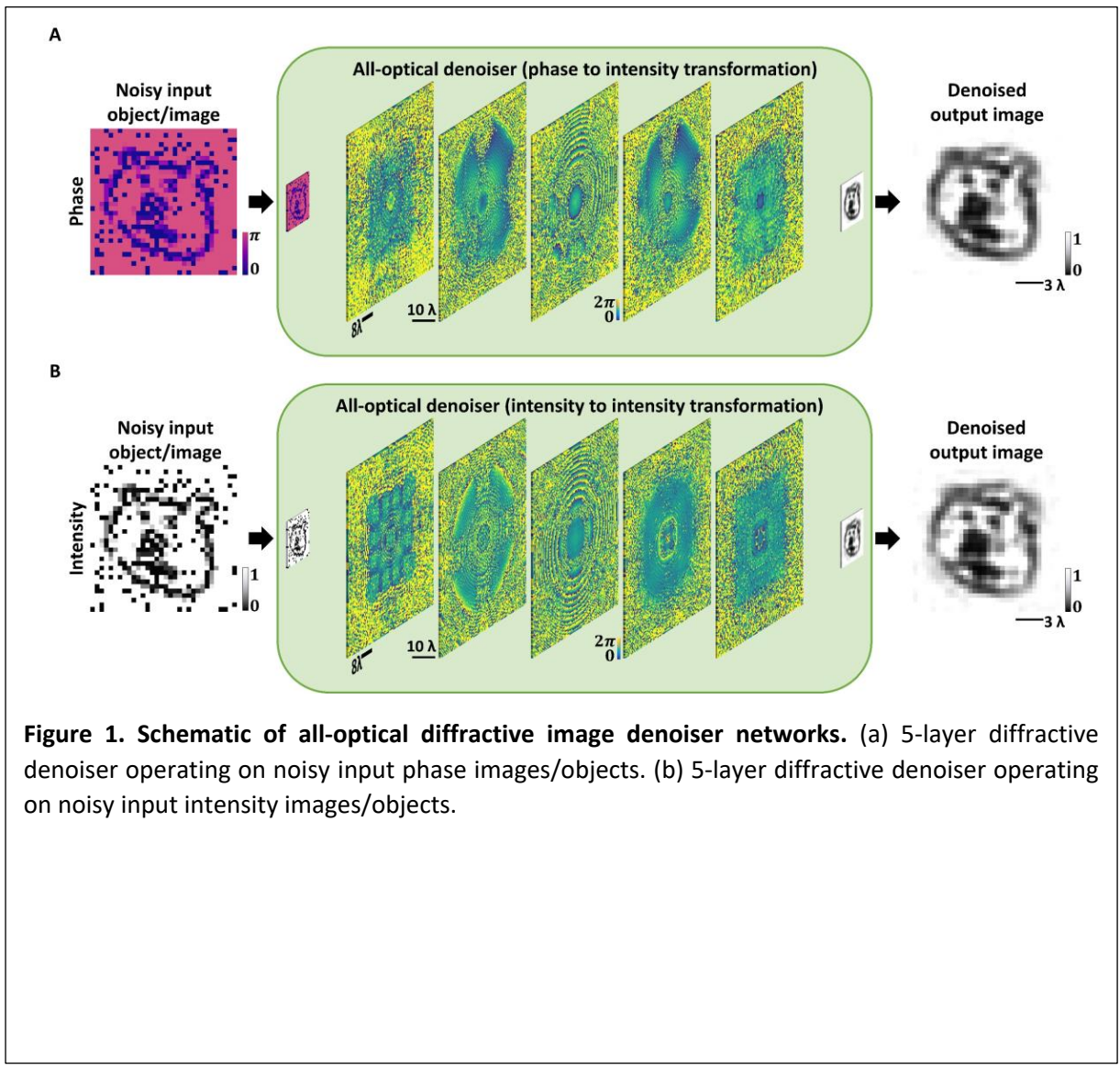

**Figure 1. Schematic of all-optical diffractive image denoiser networks.** (a) 5-layer diffractive denoiser operating on noisy input phase images/objects. (b) 5-layer diffractive denoiser operating on noisy input intensity images/objects.

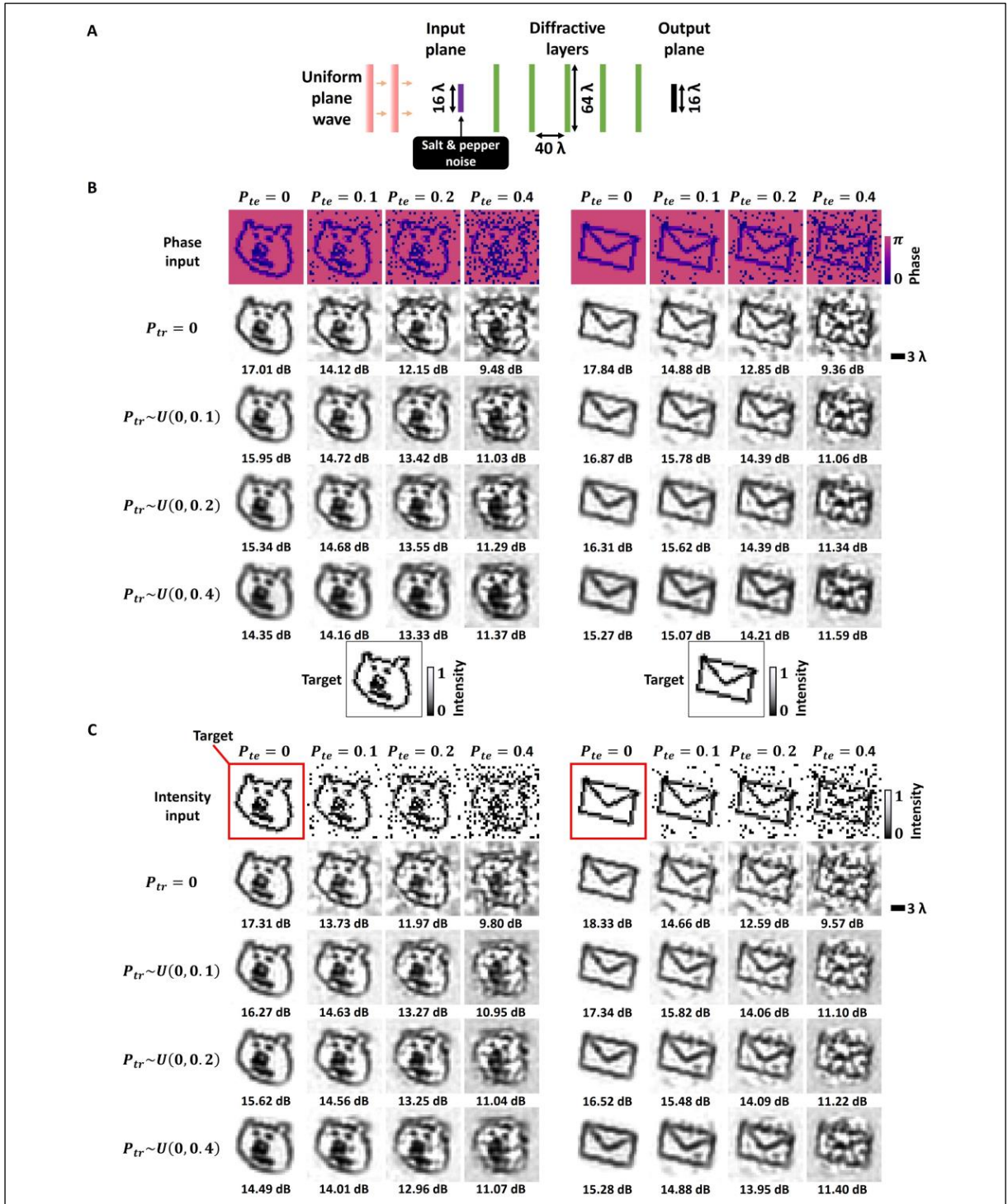

**Figure 2. Simulation results of 5-layer all-optical diffractive image denoisers for filtering out salt & pepper noise. (A)** Optical layout of the diffractive image denoisers operating on phase or intensity input images. **(B)** All-optical image denoising results of different diffractive image denoisers with phase-encoded inputs, which are trained using $P_{tr}$ drawn uniformly from different intervals. The PSNR value for each case is shown beneath the corresponding output image. **(C)** All-optical image denoising results of different diffractive denoisers with intensity-encoded inputs, trained using $P_{tr}$ drawn uniformly from different intervals.

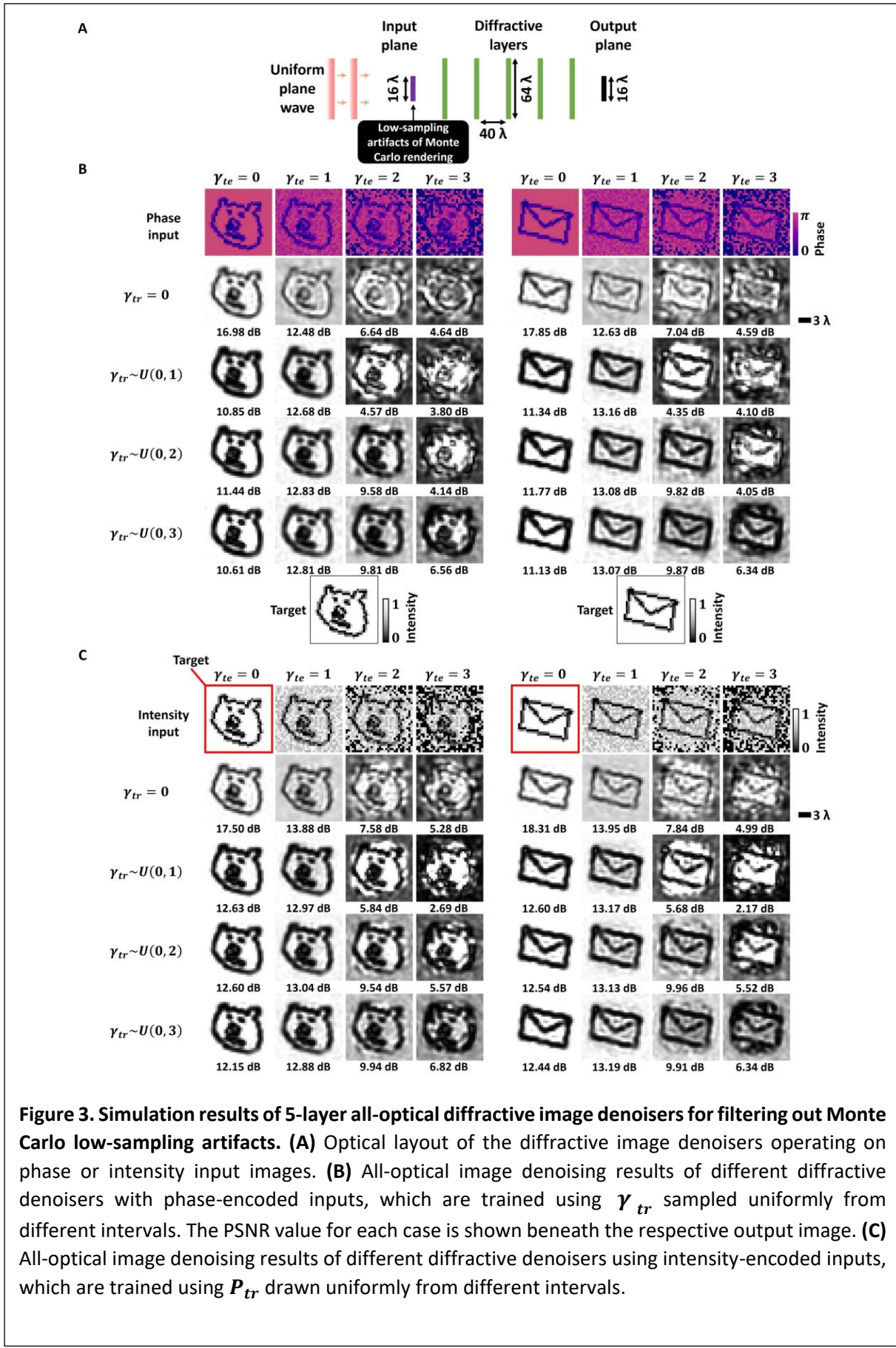

**Figure 3. Simulation results of 5-layer all-optical diffractive image denoisers for filtering out Monte Carlo low-sampling artifacts. (A)** Optical layout of the diffractive image denoisers operating on phase or intensity input images. **(B)** All-optical image denoising results of different diffractive denoisers with phase-encoded inputs, which are trained using $\gamma_{tr}$ sampled uniformly from different intervals. The PSNR value for each case is shown beneath the respective output image. **(C)** All-optical image denoising results of different diffractive denoisers using intensity-encoded inputs, which are trained using $P_{tr}$ drawn uniformly from different intervals.

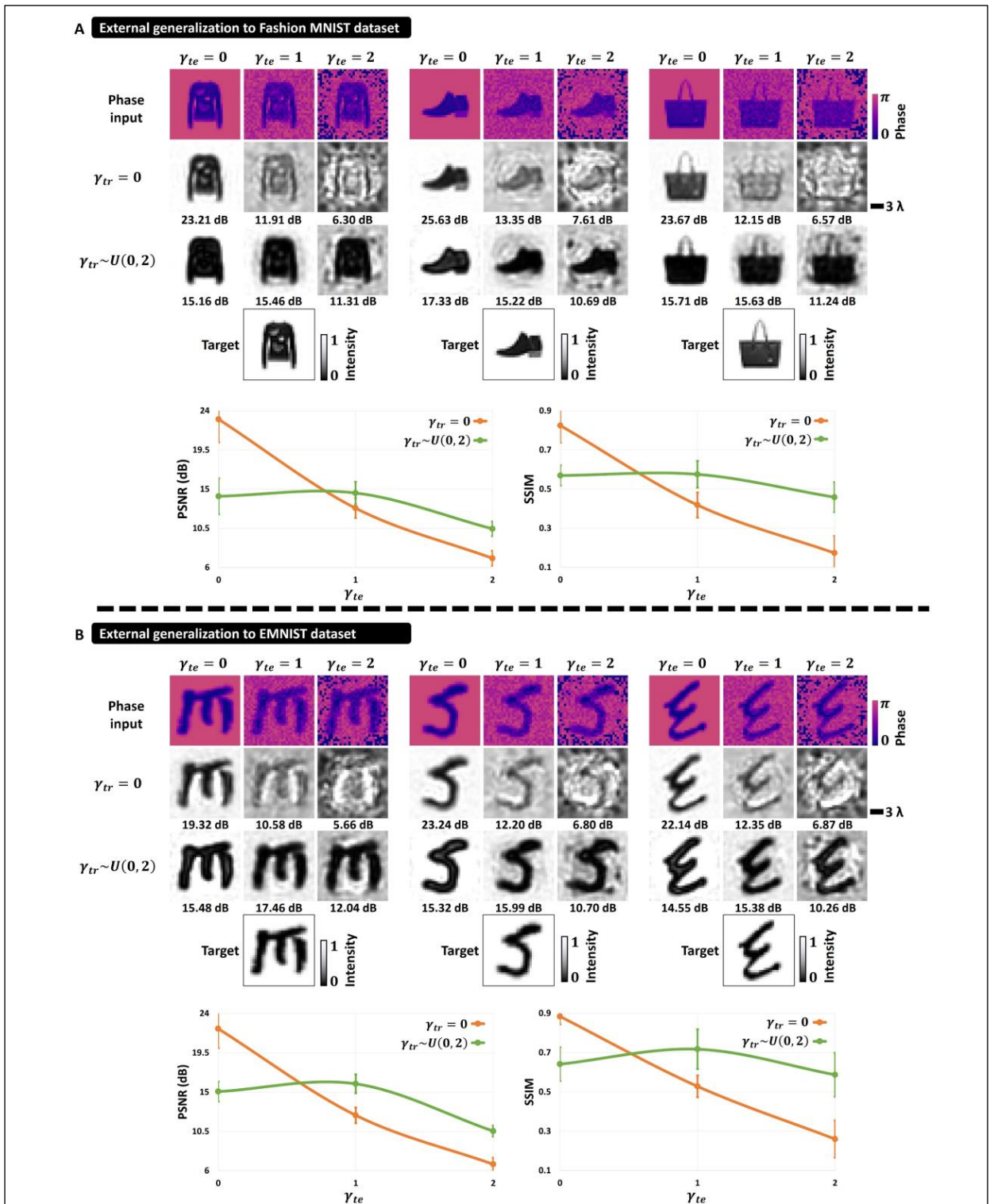

**Figure 4. External generalization performance of 5-layer all-optical diffractive image denoisers for Monte Carlo low-sampling artifact removal.** The diffractive image denoiser using phase-encoded inputs is trained with the *tiny quickdraw* dataset. **(A)** All-optical image denoising results on three images randomly selected from the Fashion MNIST test dataset and the average PSNR and SSIM values on the same test dataset as a function of $\gamma_{te}$. The PSNR value for each case is shown beneath the respective output image. **(B)** All-optical image denoising results on three images of the EMNIST test dataset and the average PSNR and SSIM values on the same dataset as a function of $\gamma_{te}$.

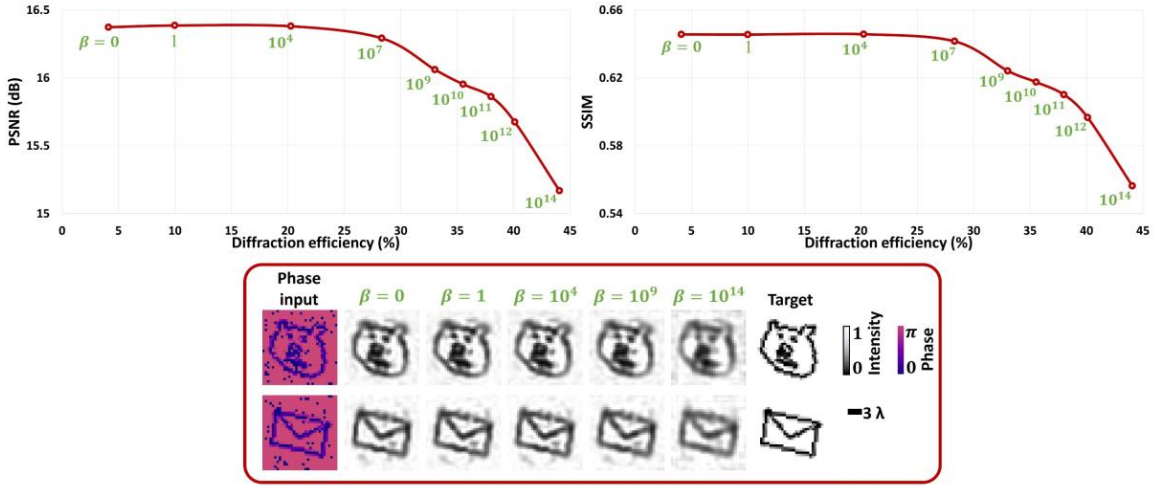

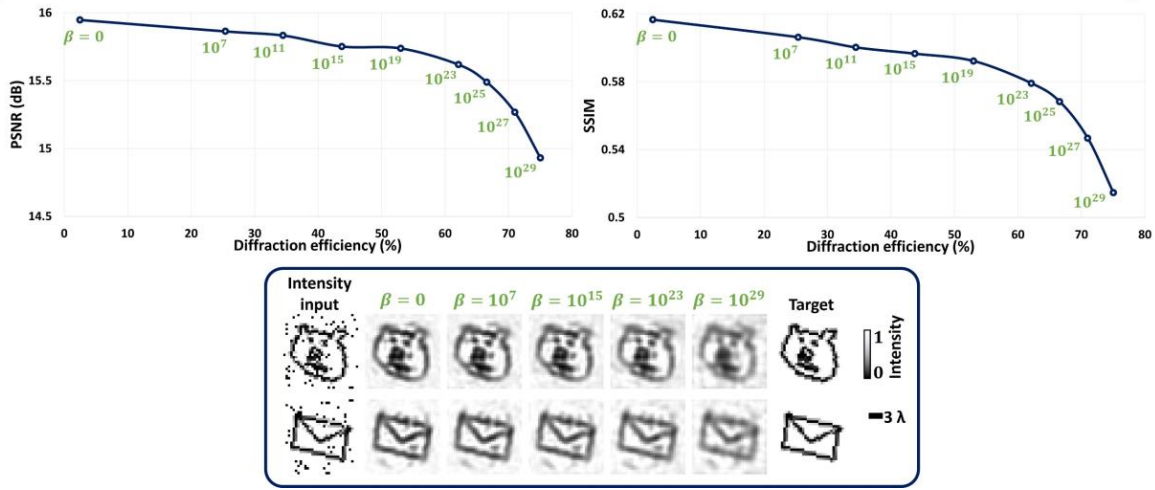

**Figure 5. Quantitative performance of 5-layer diffractive image denoisers as a function of the output diffraction efficiency for phase-encoded and intensity-encoded inputs.** The weight of the diffraction efficiency loss term ($\beta$) is varied to train the diffractive image denoisers with different output efficiencies. These all-optical image denoisers for each input type are trained using the *tiny quickdraw* training dataset under salt and pepper noise with $P_{tr}$ sampled uniformly from the interval $U(0,0.2)$. Subsequently, the trained models are tested on the *tiny quickdraw* test dataset, affected by salt and pepper noise with $P_{te} = 0.1$. **(A)** All-optical image denoising performance of the diffractive denoisers with phase-encoded inputs as a function of the average output diffraction efficiency. **(B)** All-optical image denoising performance of the diffractive denoisers with intensity-encoded inputs as a function of the average output diffraction efficiency.

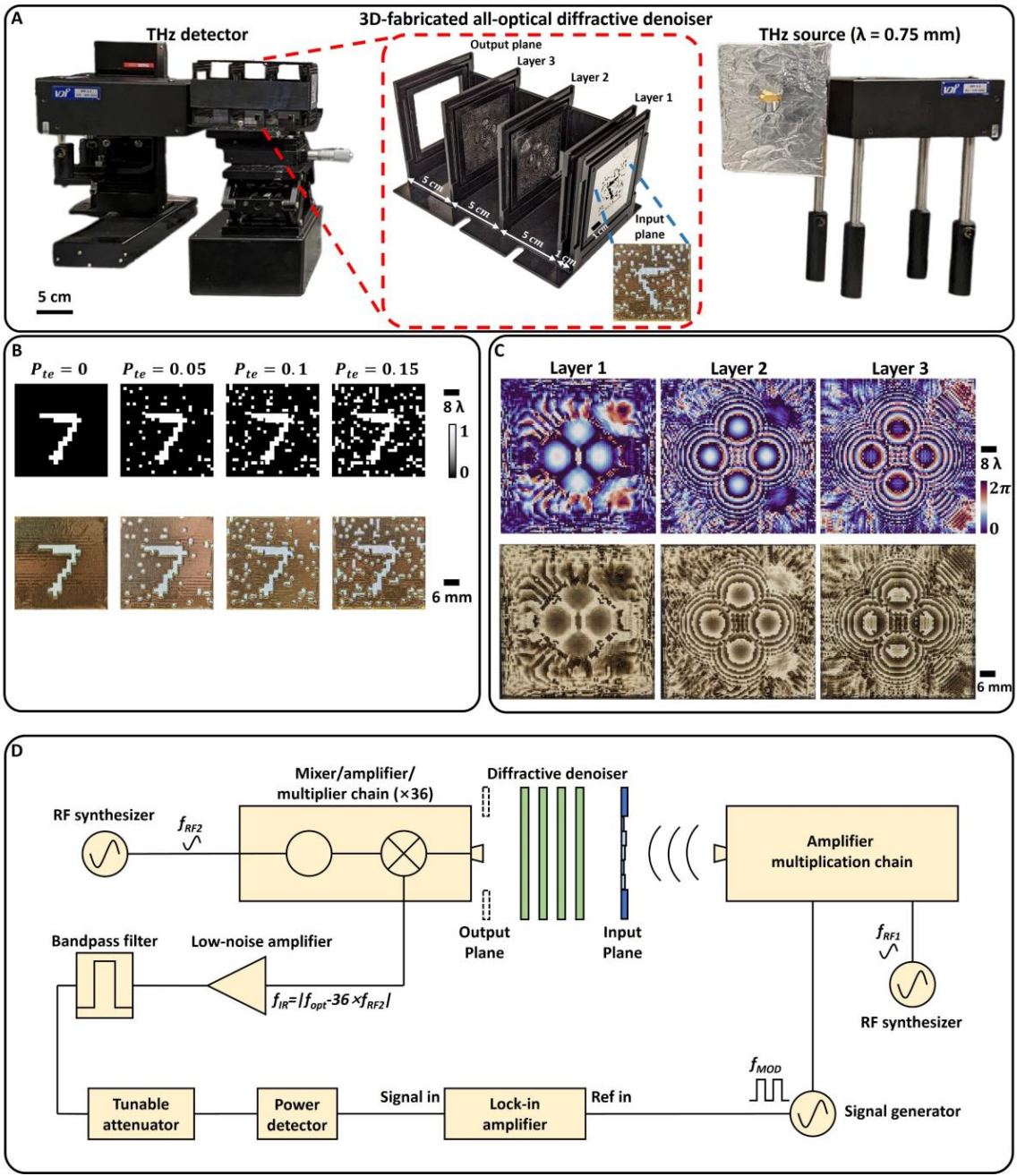

**Figure 6. Experimental setup for a 3-layer diffractive image denoiser. (A)** Photograph of the experimental setup including the 3D-fabricated all-optical image denoiser trained for noisy intensity images. **(B)** Intensity profiles of an image example impacted by various levels of salt-only noise ($P_{te}$) and their photographs after their 3D-fabrication. **(C)** Phase profiles of the trained diffractive image denoiser layers and their photographs after 3D-fabrication. **(D)** Schematic of the experimental setup using continuous wave THz illumination ($\lambda$ = ~0.75 mm).

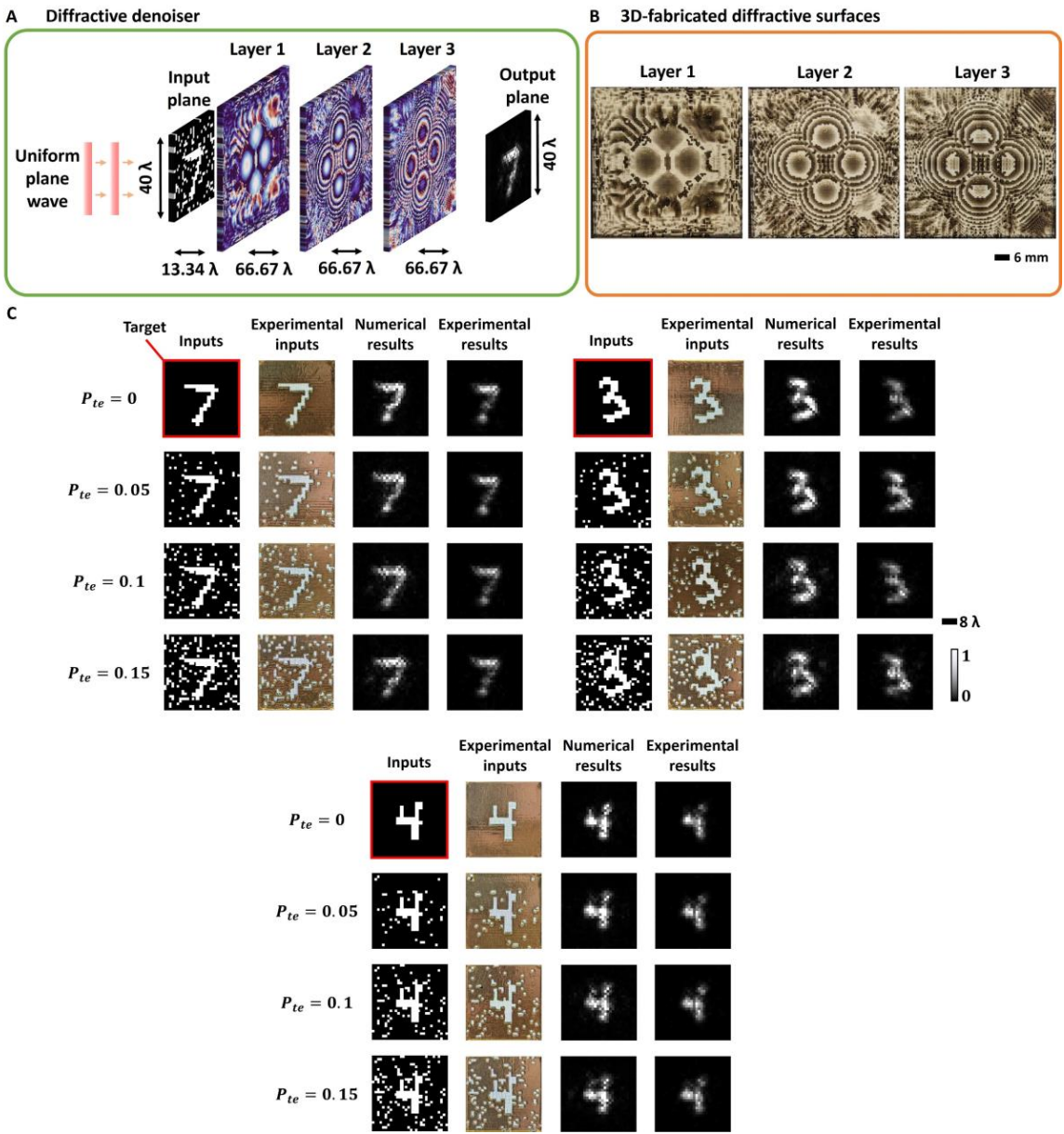

**Figure 7. Experimental results of the all-optical diffractive image denoiser. (A)** Layout of the diffractive image denoiser with 3 transmissive layers. **(B)** Photographs of 3D-fabricated layers of the trained diffractive image denoiser. **(C)** Experimental and numerical image denoising performance of the designed diffractive denoiser under different levels of salt-only noise ($P_{te}$).